\documentclass[twocolumn,preprintnumbers,amsmath,amssymb]{revtex4}
\usepackage[version=3]{mhchem} 
\usepackage{graphicx}
\usepackage{wrapfig}
\usepackage{amsmath}
\usepackage{caption}

\begin{document}

\title{Uquantchem: A versatile and easy to use Quantum Chemistry Computational Software.}

\author{Petros  Souvatzis\footnote{Email: petros.souvatsiz@fysik.uu.se}}
\affiliation{Department of Physics and Astronomy, Division of Materials Theory, Uppsala University,
Box 516, SE-75120, Uppsala, Sweden}



\begin{abstract}
In this paper we present  the Uppsala Quantum Chemistry package (UQUANTCHEM), a new and versatile computational 
platform with capabilities ranging from simple Hartree-Fock calculations to state of the art First principles Extended Lagrangian Born Oppenheimer 
Molecular Dynamics (XL-BOMD) and diffusion quantum Monte Carlo (DMC). The UQUANTCHEM package is distributed under the general public license and can be 
directly downloaded from the code web-site\cite{uquantchem}. Together with a presentation of the different capabilities 
of the uquantchem code and a more technical discussion on how these capabilities have been implemented, a presentation of  the user-friendly aspect of the package 
on the basis of the large number of default settings will also be presented.
Furthermore, since the code has been parallelized within the framework of the message passing interface (MPI),
 the timing of some benchmark calculations are reported to illustrate how the code scales with the number of computational nodes for different levels of chemical theory. 
\end{abstract}

\maketitle
\section{INTRODUCTION}
One of the main motives behind the recent development of the Uppsala Quantum Chemistry package, UQUANTCHEM)\cite{uquantchem}, has been
to  complement the broad selection of quantum chemistry codes available with an "easy to use", open source, development friendly and yet versatile computational 
framework. The other motive, which has perhaps been the most important driving force, is to provide a pedagogical platform for students and scientists  
active in the computational chemistry community that are harboring intermediate to basic programming skills, but nevertheless are interested in learning how to implement
new computational tools in quantum chemistry.  The didactical design of the code has been achieved 
by limiting the level of optimization, not to obscure the connection between  the different quantum chemical methods implemented in the package, and the actual text-book algorithms \cite{szabo,cook}, upon which the construction 
of the code rests.

The user-friendliness of the UQUANTCHEM 
package has been ascertained by a large set of default values for the computational parameters, in order for  the inexperienced user not to get overwhelmed by  technical details. 
Furthermore, thanks to the limited number of pre-installed computational libraries 
required prior to the installation of the UQUANTCHEM code (only the linear algebra package (LAPACK)\cite{lapack} and the basic linear algebra subprograms (BLAS)\cite{blas} are required) the package 
is also very simple to install.  The UQUATCHEM code has been written completely in Fortran90 and comes in three versions; A serial version, an openmp version and a MPI version. In the case of the 
serial and the openmp version,  more or less generic make files are provided for the ifortran and gfortran\cite{gfortran} compilers. The MPI version of  the UQUANTCHEM code comes with 
pre-constructed make files for five of the largest computer clusters in Sweden,  the Lindgren cluster\cite{lindgren}, the Matter Cluster\cite{matter}, the Triolith cluster\cite{triolit}, the Abisko cluster\cite{abisko}
and the Kalkyl cluster\cite{kalkyl}. These makefiles can be used as templates to create makefiles for a broad selection of clusters.

The wide range of capabilities of the UQUANTCHEM package is perhaps best illustrated by the different levels of chemical theory in which the electron correlation can be treated by UQUANTCHEM, ranging from Hartree-Fock and M{\o}ller plesset second order 
perturbation theory (MP2)\cite{MP2} to configuration interaction\cite{szabo}, density functional theory (DFT)\cite{DFT1,DFT2} and diffusion quantum Monte Carlo (DQMC)\cite{dqmc}.

The UQUANTCHEM package provides a platform on which further development can  easily be made, 
since the implementation of the different electronic structure techniques in UQUANTCHEM has, to  a large extent,  been made almost in one to one correspondence with the  text books of {\it Szabo and Ostlund} \cite{szabo} and  {\it Cook} \cite{cook}, i.e the code has been transparently written  and well commented in reference to these texts. The developer friendliness is further enhanced by the explicit calculation of all relevant data structures such as kinetic energy integrals, potential energy integrals and their 
gradients with respect to electron and nuclear coordinates. Furthermore, since the UQUANTCHEM is constructed from a very limited number of subroutines and modules, an overview of the data structure and design of the code  is easilly achieved, simplifying 
any future modification of the program.


\section{CAPABILITIES}
The UQUANTCHEM code is a versatile computational package with a number of features useful to any computational chemist. 
 The main ingredient  in any quantum chemical calculation is the level of theory in which the correlation of the electrons are treated, here the UQUANTCHEM 
package is no exception. The least computational demanding level of theory explored by the UQUANTCHEM code
is  the Hartree-Fock level of theory, where the electron correlation is completely ignored.

In the context of Hartree-Fock total energy calculations it is also possible to calculate  analytical inter-atomic forces,
  enabling the user to either relax the molecular structure with respect to the Hartree-Fock total energy,  or perform molecular dynamics (MD) calculations. Here the user can either choose to run 
a Born-Oppenheimer Molecular Dynamics calculation (BOMD), or an extended Lagrangian Molecular Dynamics calculation (XL-BOMD), where the density matrix of the next time-step 
is propagated from the previous time step by means of an auxiliary recursion relation\cite{XLBOMD}. The advantage of the XL-BOMD methodology over the BOMD approach is that in the case of XL-BOMD, there is no need of a  thermostat and an accompanying rescaling of the nuclear velocities in order to suppress any energy drift. 
\begin{figure}[tbp]
\begin{center}
  \includegraphics*[angle=0,scale=0.3]{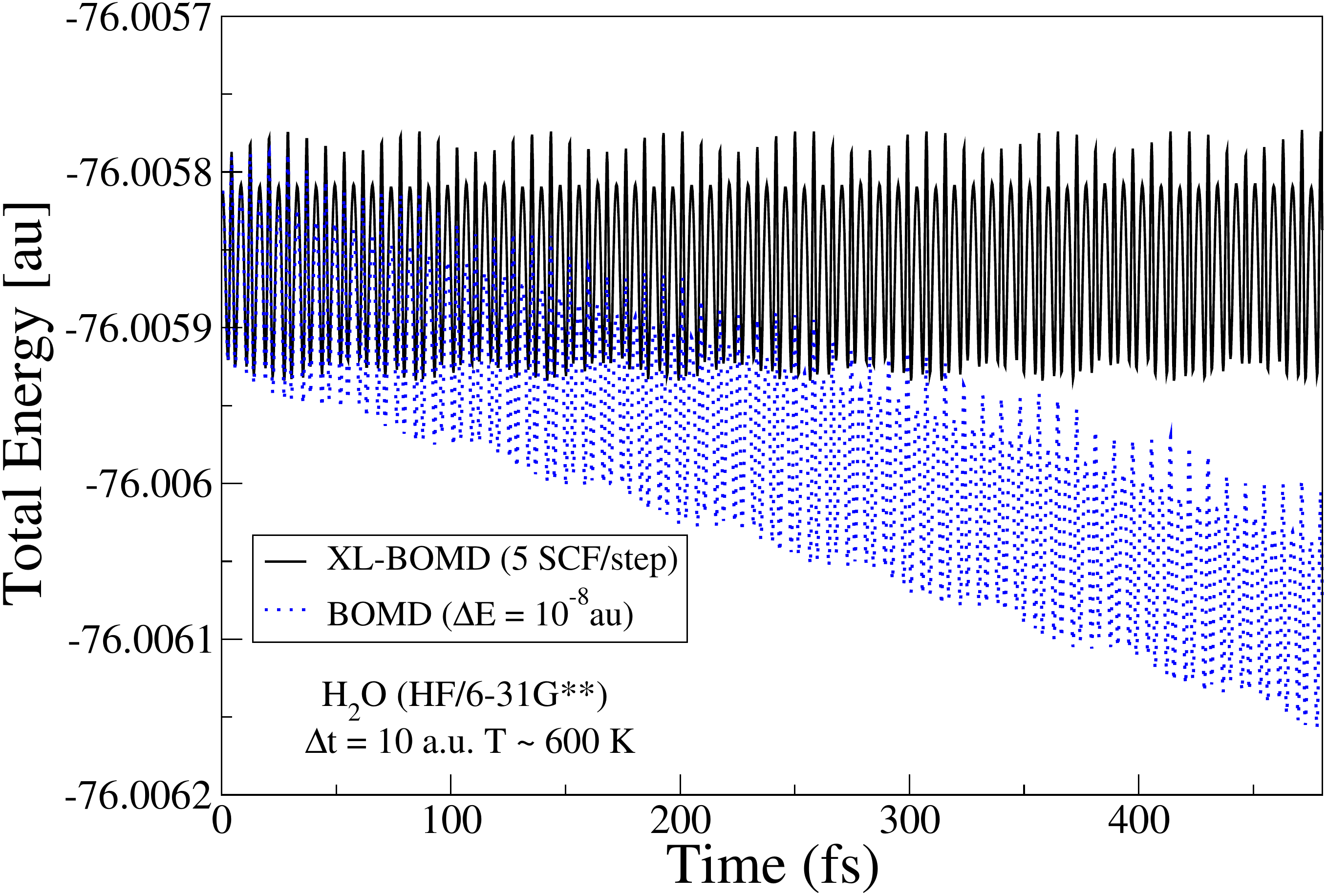}
  \caption{Total energy of  a Born-Oppenheimer molecular dynamics (BOMD) calculation of a H$_{2}$O molecule, where the density matrix
from the previous time step, is used as the initial guess to the SCF optimization with
the energy converged to $< 0.01~ \mu$Hartree, blue dotted line. The  total energy of  a Extended Lagrangian Born-Oppenheimer molecular dynamics (XL-BOMD)\cite{XLBOMD} calculation of a H$_{2}$O molecule with 5
 scf iterations per time step, full black line. }
  \label{fig:xlbomd}
  \end{center}
\end{figure}
In Figure \ref{fig:xlbomd} the results obtained with the UQUANTCHEM  code for a H$_{2}$O molecule using the  XL-BOMD  and BOMD schemes  without thermostat  are shown. Here the superiority of the XL-BOMD approach over the BOMD scheme is manifested by the lack of energy drift in the former's total energy.

On the intermediate level of electron correlation theory implemented in the package one finds MP2\cite{MP2}, CISD\cite{szabo} and DFT\cite{DFT1,DFT2}.  When using the DFT level of electron correlation it is also possible to calculate 
analytical interatomic forces, and therefore also perform structural relaxation and molecular dynamics calculations. Here the DFT forces are analytical to the extent that the gradients with respect to nuclear coordinates of the exchange correlation energy are 
calculated  as analytical gradients  of the quadrature expression used to calculate the exchange correlation energy\cite{exchforce}.

\begin{figure}[t]
\begin{center}
  \includegraphics*[angle=0,scale=0.3]{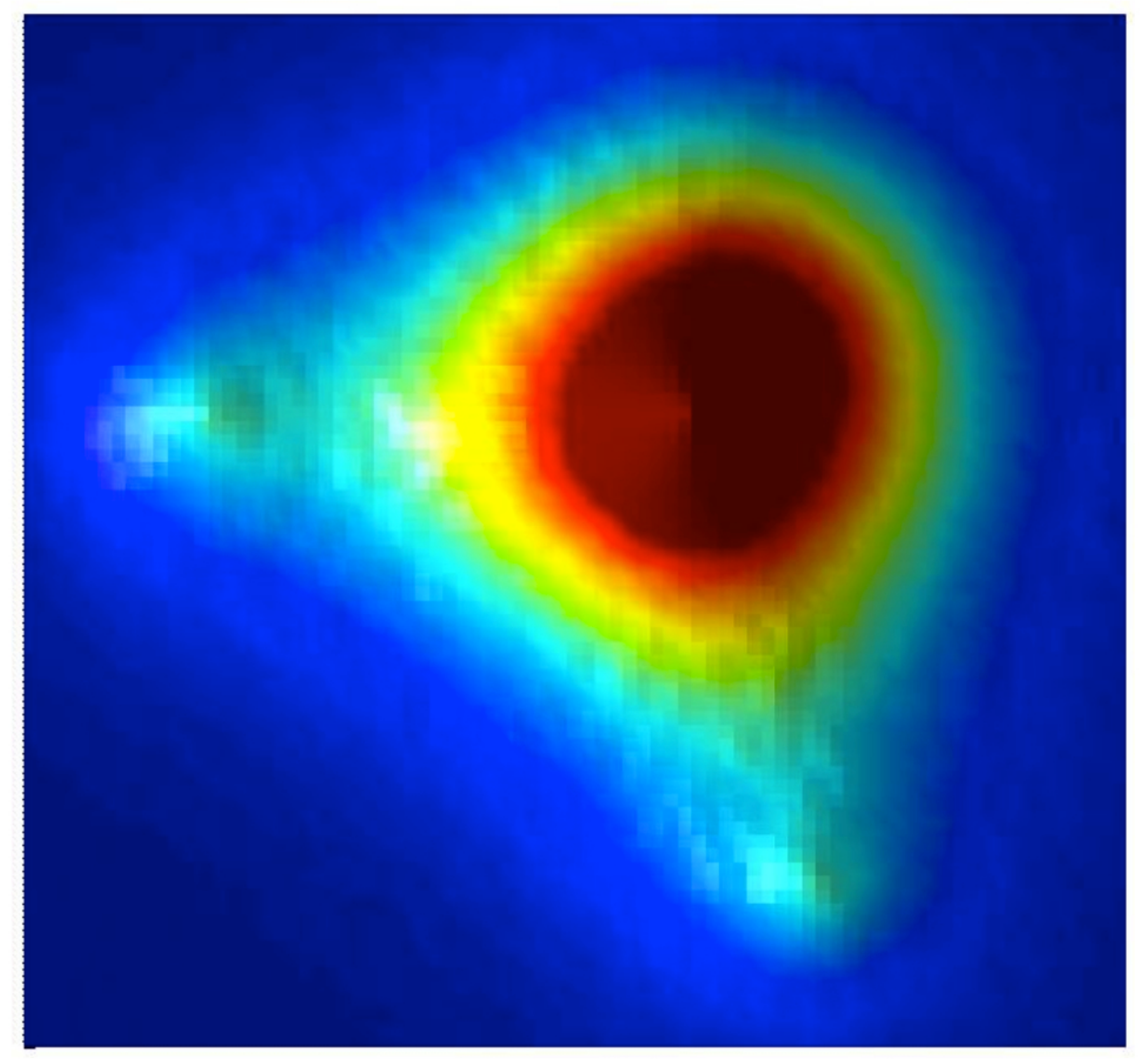}
  \caption{The estimated charge density in water using a diffusion quantum Monte Carlo algorithm as implemented in the UQUANTCHEM code. Here red colors correspond to high values of the charge density  and blue or bluish colors correspond to low values.
  The charge density was estimated from 2000 walkers and $4.0\cdot10^{6}$ time steps with a time step of $\Delta t = 1.0\cdot 10^{-4}$ au, and a cusp corrected\cite{cusp} cc-pVTZ basis. The resulting ground state energy was $E_{0}=-76.4409\pm0.0002$ a.u.
  The rendering of the charge density map was obtained by using the UQUANTCHEM output file  \texttt{CHARGEDENS.dat} as input to the Matlab script \texttt{chargdens2DIM.m}, which is a script  provided with the UQUANTCHEM package. }
  \label{fig:dqmc}
\end{center}
\end{figure}

The highest level of electron correlation theory possible to utilize within the UQUANTCHEM package is DQMC\cite{dqmc}. Here it is possible, within the fixed node approximation\cite{fixnode}, to calculate 
total ground state energies of medium sized molecules taking into account $>90\%$ of the correlation energy. In Figure \ref{fig:dqmc}  the estimated charge density of a  H$_{2}$O molecule is shown, here calculated with DQMC as implemented in 
UQUANTCHEM.

Apart from the different methods involved in dealing with electron correlation implemented in the UQUANTCHEM package, a number of other capabilities can also be found within the package. Amongst these capabilities is the ability 
to  provide graphical information about the highest occupied and the lowest unoccupied orbitals (HOMO and LUMO),  deal with charged systems as well as calculating Mulliken charges,  plot the Hartree-Fock and Kohn-Sham orbitals as well as the corresponding charge density, calculate the velocity auto-correlation function and relax the molecular structure with respect to interatomic forces by utilizing a conjugate gradient scheme.
 However, it should be stressed that the code can only deal with finite systems, thus excluding any calculation of periodic systems.
									
\section{IMPLEMENTATION DETAILS}
The code utilizes a localized atomic basis set, where each basis function, $\phi_{i}$,  is constructed from a contraction of primitive cubic Gaussian orbitals\cite{cook}
 \begin{align}
 \phi_{i}(x,y,z)  = \qquad  \qquad  \qquad  \qquad  \qquad  \qquad  \qquad \qquad \quad  \nonumber  \\
 \nonumber  \\
  \sum_{n}d_{in}(x-X_{i})^{L_{i}}(y-Y_{i})^{M_{i}}(z-Z_{i})^{N_{i}}e^{-\alpha_{i}(\mathbf{r}-\mathbf{R}_{i})^2}. 
\end{align}
 Here, $d_{in}$ are the contraction coefficients, $\mathbf{R}_{i}=(X_{i},Y_{i},Z_{i})$ are the atomic coordinates at which the basis function is centered, $\alpha_{i}$ are the primitive Gaussian exponents and $L_{i}, M_{i}$ and $N_{i}$ integer 
 numbers determining the angular momentum, $l = L_{i} + M_{i} + N_{i}$, of the corresponding basis function. In what follows we will suppress the spin part of the basis functions and assume that the spin degrees of freedom are treated 
 implicitly, i.e have been integrated out.
 
Thanks to the use of primitive Gaussians almost all the matrices involved in the different implementations, such as the overlap matrix, $S$, the kinetic energy matrix, $T$ and the nuclear attraction matrix, $V$ defined by
 \begin{eqnarray}
 S_{ij} = \int d^{3}\mathbf{r}\phi_{i}(\mathbf{r})\phi_{j}(\mathbf{r}), \\
 T_{ij} = -\frac{1}{2}\int d^{3}\mathbf{r}\phi_{i}(\mathbf{r})\nabla^{2}\phi_{j}(\mathbf{r}),\\
 V_{ij} = -\sum_{n}Z_{n}\int  d^{3}\mathbf{r}\frac{\phi_{i}(\mathbf{r})\phi_{j}(\mathbf{r})}{|\mathbf{r}-\mathbf{R}_{n}|}, 
 \end{eqnarray}
have been calculated analytically. Here, $Z_{n}$ denotes the atomic numbers of the nuclei. The implementation of the analytic evaluation of the above integrals follows almost exactly the outline given in D. B. Cook's book {\it Handbook of Computational Chemistry}.\cite{cook} 

In order to enhance the performance of the code, the electron-electron integrals,
\begin{equation}
(ij|kl) = \int d^{3}\mathbf{r}\int d^{3}\mathbf{r'}\frac{\phi_{i}(\mathbf{r})\phi_{j}(\mathbf{r})\phi_{k}(\mathbf{r'})\phi_{l}(\mathbf{r'})}{|\mathbf{r}-\mathbf{r'}|},
\end{equation}
have been calculated by  Rys quadrature\cite{rys}, even though they can be calculated analytically as is described in Cook's book\cite{cook}.
 
The exchange correlation energy, $E_{xc}$ and the corresponding exchange correlation matrix elements, $V^{xc}_{ij}$ 
 \begin{eqnarray}
E_{xc} = \int d^{3}\mathbf{r}\epsilon_{xc} [ \rho,\nabla \rho ]\rho(\mathbf{r}), \qquad  \qquad \qquad \\
V^{xc}_{ij} =  \int d^{3}\mathbf{r}\phi_{i}(\mathbf{r})\frac{\delta E_{xc}}{\delta \rho}\phi_{j}(\mathbf{r})=   \nonumber \qquad \qquad  \\ 
=\int d^{3}\mathbf{r}\phi_{i}(\mathbf{r})\frac{d (\epsilon_{xc}\rho)}{d \rho}\phi_{j}(\mathbf{r}) + \tilde{V}^{xc}_{ij},  \qquad \qquad \\
\tilde{V}^{xc}_{ij} = \int \frac{d^{3}\mathbf{r}}{|\nabla\rho|}\Big ( \nabla \phi_{i}\cdot\nabla\rho \phi_{j} + \phi_{i}\nabla\rho\cdot \nabla \phi_{j}\Big )\frac{d(\epsilon_{xc}\rho)}{d |\nabla\rho|}
\end{eqnarray}
defined through the exchange correlation energy density, $\epsilon_{xc}[\rho]$, and the  functional derivative of the exchange correlation energy \cite{Correctfder}, 
\begin{eqnarray}
V_{xc}(\mathbf{r}) = \frac{\delta E_{xc}}{\delta \rho} \equiv \nonumber  \qquad \qquad \qquad \qquad  \qquad  \qquad  \qquad  \qquad     \quad \\
\frac{d( \epsilon_{xc}\rho)}{d \rho}-\frac{d}{dx}\Big(\frac{d( \epsilon_{xc}\rho)}{d (\nabla\rho)_{x}}\Big)-\frac{d}{dy}\Big(\frac{d( \epsilon_{xc}\rho)}{d (\nabla\rho)_{y}}\Big)-\frac{d}{dz}\Big(\frac{d( \epsilon_{xc}\rho)}{d (\nabla\rho)_{z}}\Big), \nonumber
\end{eqnarray}
have been calculated by decomposing  the above spatial integrals
 into sums of integrals over atom-centered "fuzzy" polyhedra, as described by Becke\cite{GRID}. Each one of these polyhedral integrals 
is computed by the use of Gauss-Chebyshev quadrature of second order\cite{GAUSSCHEB}, and Lebedev quadrature\cite{LEB}. Here, $\rho$ denotes the charge density. 

The only difference between the implementation of the above integrals in UQUANTCHEM and the outline given by Becke\cite{GRID}, is that in the UQUANTCHEM implementation, the mapping of the interval $x\in[-1,1]$ 
into the radial integration interval $r\in[0,\infty]$, enabling the use of Gauss-Chebyshev quadrature, is  contrary to what is prescribed by Becke done with the mapping
\begin{equation}
r = -r_{m}log \Big [ 1 - \Big ( \frac{x+1}{2} \Big )^{4} \Big ].
\end{equation}
Here $r_{m}$ is the Slater atomic covalent radius\cite{slaterr} of the atom at which the corresponding "fuzzy" tetrahedron is centered. It has been shown by Mura {\it et al}\cite{RGRID} that the above mapping 
results in a much more accurate numerical integration as compared to what is achieved with the mapping proposed by Becke\cite{GRID}.

The density functionals provided with the UQUANTCHEM package are the local density approximation (LDA) functional of Vosko {\it et al}\cite{VWN}, the revised gradient corrected functional of Perdew {\it et al}\cite{PBE} (revPBE)\cite{PBErev} and the 
B3LYP hybrid functional\cite{becke88,VWN,LYP,b3lyp}. 

The MP2 implementation and the CISD implementation very closely follow the outline given in the text-book of Szabo and Ostlund \cite{szabo}.

 The implementation of the DQMC algorithm  in UQUANTCHEM follows 
closely the algorithm outlined in the work of Umrigar {\it et al} \cite{dqmc}. However, in UQUANTCHEM the trial function is constructed 
with a much simpler Jastrow factor, $\mathcal{J}$,  and the slater determinants are constructed from cusp corrected Gaussian orbitals, instead of Slater type orbitals (STO), as in the work of  Umrigar {\it et al}. The implementation of the cusp correction 
in UQUANTCHEM follows the prescription given by S. Manten and A. L\"uchow \cite{cusp}. 
The explicit form of the trial function used for the importance sampling in the DQMC of UQUANTCHEM is given by:
\begin{equation}
\Psi_{T} = D_{\uparrow}D_{\downarrow}\mathcal{J}.
\end{equation}
Where 
\begin{equation}
\mathcal{J} = exp\big(\sum_{i<j}\frac{\delta\cdot b \cdot r_{ij}}{(1+ c\cdot r_{ij})}\big),
\end{equation}
$r_{ij} = | r_{i} - r_{j}|$ is the distance between electron $i$ and $j$, $D_{\uparrow}$ and $D_{\downarrow}$ are the slater determinants created from spin up respectively 
spin down orbitals. The orbitals are constructed from the unrestricted Hartree-Fock (URHF) self consistent solution. Here $\delta = 0.25$ if the spin of the electrons $i$ and $j$ are identical otherwise 
if the spins are opposite, $\delta = 0.5$. Here, $b$ and $c$ are Jastrow parameters that can be chosen and optimized by the user.

\section{TECHNICAL DETAILS}
As was mentioned in the introduction, the UQUANTCHEM package has been written with the aim of keeping a high level of transparency in order to facilitate further development,  with the result  of a somewhat limited computational speed for the serial implementation of the code. A quantitative illustration of the performance limitation of the UQUANTCHEM code is obtained by comparing the total execution times of  the UQUANTCHEM code with corresponding times of  the GAMESS code \cite{gamess}. When comparing the execution times for  a HF total energy calculation employing a cc-pVTZ basis set,  the GAMESS code is about 15 times faster on a dual intel core processor. And if one instead would compare the performance of the two codes when doing a B3LYP 
calculation, on the same system, with the same basis and the same machine,  the GAMESS code come out to be $\sim$ 150 times faster.
Therefore we will in this section more carefully  discuss these performance limitations and how they  have  been dealt with by means of parallelization within the context of the message passing interface (MPI).  

\begin{figure*}[tbp]
\begin{center}
  \includegraphics[angle=0,scale=0.3]{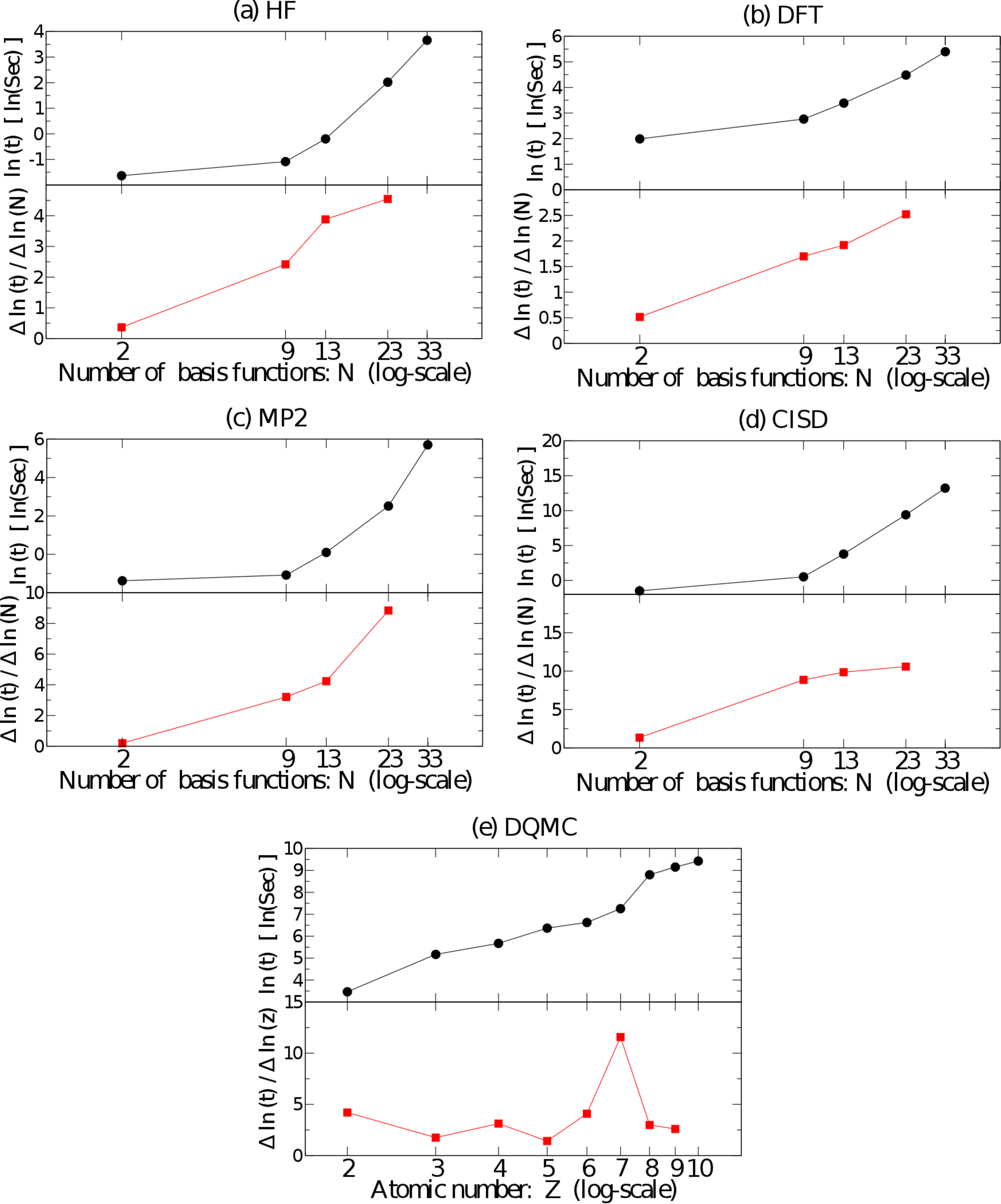}
  \caption{Computational time as a function of number of basis functions in (a-d), and number of electrons in (e), for the different levels of theory implemented in the UQUANTCHEM code. In all panels (a-e) the logarithm of the computational times are represented by black circles in the upper sub-panels  and  the finite difference, $\frac{\Delta ln(t)}{\Delta ln(N)}$ ( in (e) $\frac{\Delta ln(t)}{\Delta ln(Z)}$) , as red squares in the lower sub-panels. In (a) the computational times and finite differences of the Hartree-Fock implementation, in (b) the computational times and finite differences of  the DFT (revPBE functional) implementation, in 
 ({c}) the computational times and finite differences of the MP2 implementation, in (d) the computational times and finite differences of the CISD implementation and in (e) the computational times and finite differences of the DQMC implementation. The 
 calculations presented in panels (a-d) were performed for the atoms with atomic number $Z=2,10,18,36,54$. In all calculations a 3-21G basis set was used except for the DQMC calculations where the cc-pVTZ basis-set was employed. }
  \label{fig:time}
  \end{center}
\end{figure*}

In Figure \ref{fig:time} (a-e), in order to highlight the performance of the code, the results of a series of total energy calculations obtained with the serial version of the UQUANTCHEM code are displayed.
Here in (a-d), the logarithm of the computational time, $ln(t)$, and the finite difference, $\Delta ln(t)/\Delta ln(N)$, as a function of number of basis-set functions are displayed,  
and in (e), the logarithm of the computational time, $ln(t)$, and the finite difference, $\Delta ln(t)/\Delta ln(Z)$, as a function of number of electrons Z are displayed. From these finite 
differences the computational time of the HF implementation can be seen to scale as $\mathcal{O}(N^{4.5})$, and the computational time of the DFT implementation scale as $\mathcal{O}(N^{2.5})$, which 
are comparable with the nominal scaling of these methods, $\mathcal{O}(N^{4})$ and $\mathcal{O}(N^{3})$, respectively, found in the literature. However, when it comes to the computational time scaling of the MP2 and CISD implementation in 
UQUANTCHEM, with nominal scalings of  $\mathcal{O}(N^{5})$ and $\mathcal{O}(N^{6})$, respectively,  the situation is worse. Here the computational time scales as $\mathcal{O}(N^{8.8})$ for a basis set size of $N > 23$, for the MP2 implementation, and as $\mathcal{O}(N^{10})$ for the CISD
implementation for basis sets of size $N > 9$. Finally, from the lower panel of figure \ref{fig:time} (e) the computational time of the DQMC calculation can be seen to scale approximately as $\mathcal{O}(N^{3})$, which is basically equal to 
the nominal scaling found in the literature. Here the jump in the finite difference at Z = 3, in figure \ref{fig:time} (e), is related to the stochastical nature of the DQMC scheme.

\begin{figure}[tbp]
\begin{center}
  \includegraphics*[angle=0,scale=0.3]{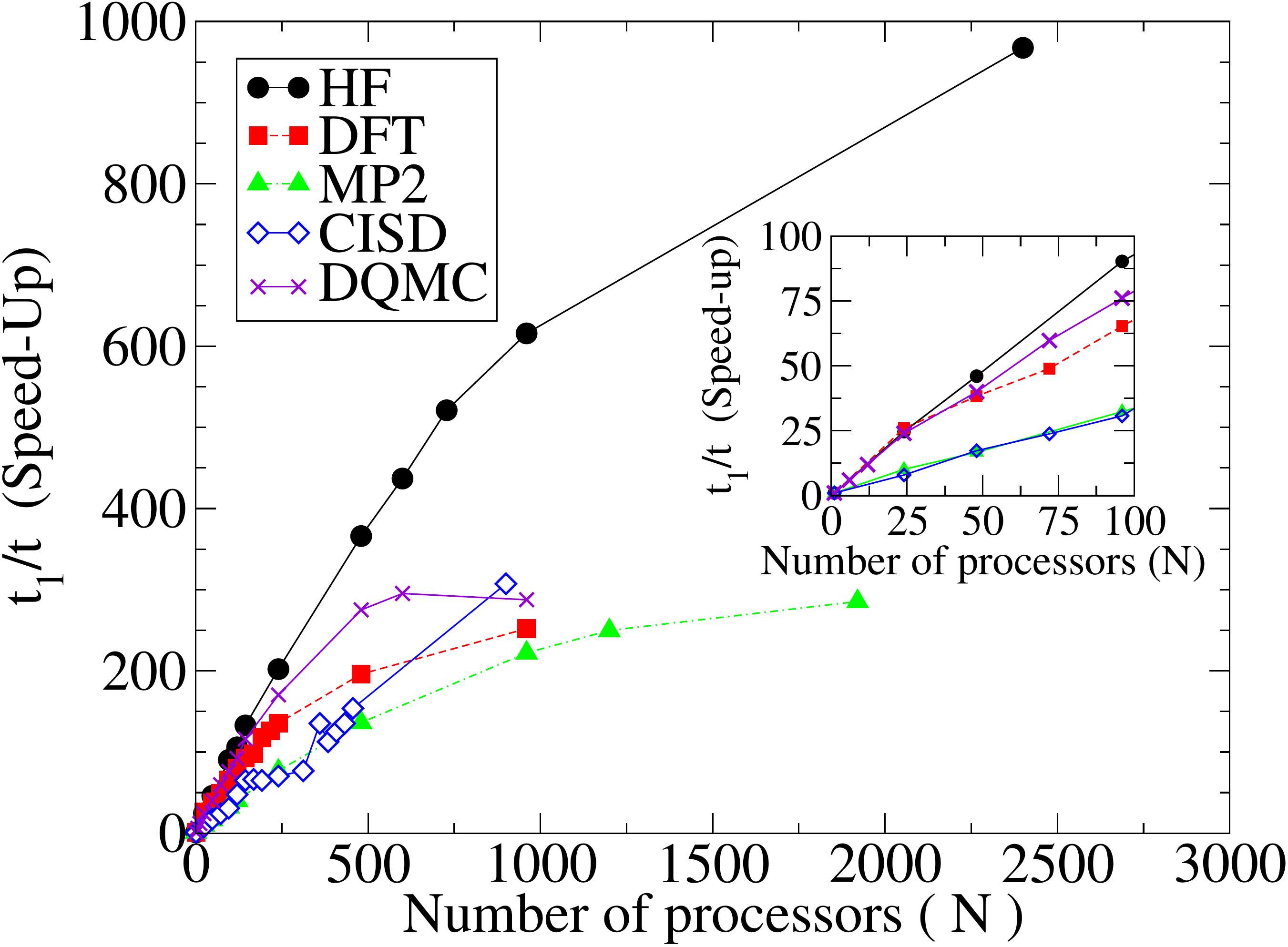}
  \caption{Speed-up as a function of the number of processors used for calculations performed at different levels of theory.
  Here the black circles is the speed-up of a Hartree-Fock calculation of H$_{2}$O using a cc-pVQZ basis set, the red squares is the speed-up 
   of a density functional theory (DFT) \cite{DFT1,DFT2} calculation of H$_{2}$O using the Perdew, Burke and Ernzerhof  (PBE) functional\cite{PBE} and a cc-pVTZ basis set , the green triangles
  is the speed-up of a second order M\"oller Plesset (MP2) \cite{MP2} calculation of H$_{2}$O using a cc-pVTZ basis set, the empty blue diamonds is the speed up of a singles and doubles configuration interaction (CISD) calculation of H$_{2}$ 
  using a cc-pVTZ basis set and the violet crosses is the speed-up of a quantum diffusion Monte Carlo (QDMC) \cite{dqmc} calculation of a single Be atom utilizing a cc-pV5Z basis set. In all the calculations. Here the computational time on one single processor, $t_{1}$,was the following:
  $t_{1}$ = 13547 s for the Hartree-Fock calculation,
  $t_{1}$ = 1764 s, for the DFT calculation,  $t_{1}$ = 5995 s for the MP2 calculation,  $t_{1}$ = 3381 s for the CISD calculation and $t_{1}$ = 44300 s for the QDMC calculation. The inset  shows an enlarged  portion of the figure  in order  to more clearly expose the 
region in which the different paralellizations are close to maximum efficiency.}
  \label{fig:timing}
  \end{center}
\end{figure}

In Figure \ref{fig:timing} the speed-up  
relative to the serial execution time, $t_{1}$,  for a couple of total energy benchmark calculations of the UQUANTCHEM  MPI version is  displayed.
Here the most effective paralellization, when using around 500 processors,  is found  
in the Hartree-Fock (HF) and diffusion quantum Monte Carlo (DQMC) implementations, where the speed-ups of $\sim 380$, in the case of the HF implementation, and $\sim 280$, in the case of the DQMC implementation, correspond to efficiencies of the respective 
parallelizations of 76\% respectively 56\%. The speed-up
of the MP2 and CISD implementations, at the same number of processors, only reaches $\sim$ 30\% efficiency, with the DFT paralellization performing slightly better. When using more than 1000 processors 
 the HF level of theory prevails as the most effectively parallelized method, at which point the speed-up of  DQMC method has already saturated and been overtaken by the CISD method. Before we continue, 
 we note in passing that the number of processors at which the speed-up (at any level of theory) is saturated strongly depends on the serial execution time, $t_{1}$, i.e on the system size and the number of 
 basis functions used in the calculation. Therefore it should be stressed that it is more informative to compare the efficiency of the different parallelizations rather than the number of processors at which the speed-up saturates.

The difference in efficiency between parallel HF calculations and the parallel MP2 and CISD calculations, comes from the fact that in  the case of the HF calculation, only the computation of two-electron integrals,
 $(ij|kl)$, and the contraction of these integrals with the density matrices $P_{ij}$ into Fock, $J_{ij}$, and exchange, $K_{ij}$, matrices , i.e
 \begin{equation}
 J_{ij} = \sum_{kl}(ij|kl)P_{kl}\quad,\quad  K_{il} = \sum_{jk}(ij|kl)P_{jk},
 \end{equation}
are parallelized. Whereas in the case of the MP2 calculation, also the sum of two-particle excitations\cite{szabo}  are made in parallel, and in the case of the CISD calculations, both the construction 
of the Hamiltonian and the diagonalization are made in parallel, thanks to the utilization of the Scalable Linear Algebra Package (scaLAPACK)\cite{scalapack} {\it divide and conquer} diagonalization routine \texttt{PDSYEVD}\cite{pdysevd}.
Furthermore, since the storage of the CISD Hamiltonian is shared amongst all the computational nodes taking part in the calculation, memory bottlenecks can be avoided, permitting the computation and diagonalization  of Hamiltonians of 
substantial size. 

The difference in the performance between parallel implementations of   HF and DFT comes from the fact that in the DFT implementation, not only the computation of two-electron integrals and their contraction are parallelized, 
as in the HF implementation, but also the calculation of the exchange-correlation potential and exchange correlation energy by means of numerical quadrature is parallelized.

Furthermore, in order to accelerate the convergence of the HF and DFT self consistent calculations,  the direct inversion of the iterative subspace algorithm (DIIS)  of P. Pulay\cite{pulay1,pulay2} 
has also been implemented in both the serial and parallel versions of UQUANTCHEM.
\section{USING THE CODE}
The UQANTCHEM package has been developed to be  easily installed and run on UNIX-type platforms. The code has been compiled without any problems on machines 
running under  Mac OSX 10.7, Red-Hat Linux and Ubuntu, using either the gfortran or ifort compilers.

The interface of the code with the user goes mainly through one single file, the \texttt{INPUTFILE}-file, through which the user specifies what type of calculation is going to be performed 
on what type of molecule. Below, the \texttt{INPUTFILE}-file specifying a DFT calculation of the total energy and HOMO-LUMO orbitals of a water molecule is given. \\ \\
\texttt{CORRLEVEL B3LYP \\
TOL 1.0E-8 \\
WHOMOLUMO .T. \\
Ne 10 \\
NATOMS 3 \\
ATOM 1~~0.4535 1.7512 0.0000 \\
ATOM  8~~0.0000 0.0000 0.0000 \\
ATOM 1 -1.8090 0.0000 0.0000}\\ \\
Apart from the \texttt{INPUTFILE}-file the user is only required to provide the code with the so called \texttt{BASISFILE}-file, specifying which type of Gaussian basis set to use in the calculation. The most 
common basis-sets used in modern quantum chemistry are provided within the UQUANTCHEM package. There is also the possibility to use the information at the {\it Basis Exchange Portal}, located at \texttt{https://bse.pnl.gov/bse/portal}, 
to extend the default library of basis-sets that comes with the UQUANTCHEM package.

Once the user has made sure that the files \texttt{INPUTFILE} and \texttt{BASISFILE} are located in the same directory as the code is being executed,  then,  in order to run for instance the serial version of the code, it is just a matter of giving 
the command:  \\
 "\texttt{commad-prompt>./uquantchem.s}"\\
 on the command line.
 
 Thus from the above given expos\'{e}, it becomes apparent that the user can, after the code has been compiled, quickly start performing quantum chemical calculations without getting overwhelmed by a plethora 
 of input-parameters and input-files. However, when necessary, the user can always bypass the default settings and more carefully specify the calculation with the computational input-parameters that were initially 
 "hidden".
\section{BENCHMARKING OF DFT}
In Table \ref{table:table1} and Table \ref{table:table2}, we show the results of a series of geometry optimization calculations obtained with Uquantchem and the well established Gamess code  \cite{gamess}. Here the results obtained with Uquantchem are virtually  indistinguishable from the results obtained with the Gamess code, especially when comparing the results obtained with the B3LYP hybrid functional. In order to calculate the exchange correlation potential  matrix elements in these Uquantchem calculations,  Gauss-Chebyshev quadrature of the second kind with 100 grid points was used for the radial integration, and Lebedev quadrature \cite{LEB} using  194 angular grid points was used for the angular part of the integration. The structures of the molecules were considered optimized when the absolute values of the interatomic forces had been relaxed down to  $<10^{-6}$ a.u.

In Table \ref{table:table3}, the corresponding experimental structural parameters are also displayed.

\begin{table*}[htb]
\footnotesize
\centering
\captionsetup{width=0.85\textwidth}
\caption{Results from geometry optimization using the revPBE \cite{PBErev} functional,  with 6-31G**. Here, $\alpha$, is the equilibrium angle, d, in units of [\AA], the equilibrium distance between nearest neighbor atoms and, E, the total energy for the optimized structure.}
 \begin{tabular}{c | c c c | c c c | c c c}
\hline\hline
 & \multicolumn{3}{c}{H$_{2}$O} & \multicolumn{3}{c}{NH$_{3}$} & \multicolumn{3}{c}{CH$_{4}$} \\
\hline
  \hline
  Code & $\alpha$(HOH) & d(OH) & E [a.u.] & $\alpha$(HNH) & d(NH)  & E [a.u.] & $\alpha$(HCH) & d(CH) & E [a.u.] \\
\hline
Gamess         &  102.571 & 0.9736 & -76.3940  & 104.495 & 1.0271 & -56.5378 & 109.471 & 1.0996 & -40.5050\\
Uquantchem &  102.598 & 0.9734 & -76.3956  & 104.508 & 1.0269 & -56.5391 & 109.469 & 1.0994 & -40.5059\\
\hline
\end{tabular}
\label{table:table1}
\end{table*}
\begin{table*}[htb]
\footnotesize
\centering
\captionsetup{width=0.85\textwidth}
\caption{Results from geometry optimization using the B3LYP functional  \cite{becke88,VWN,LYP,b3lyp}, with 6-31G**. Here $\alpha$ is the equilibrium angle, d,  in units of [\AA],  the equilibrium distance between nearest neighbor atoms and, E, the total energy for the optimized structure.}
\begin{tabular}{c | c c c | c c c | c c c}
\hline\hline
 & \multicolumn{3}{c}{H$_{2}$O} & \multicolumn{3}{c}{NH$_{3}$} & \multicolumn{3}{c}{CH$_{4}$} \\
\hline
  \hline
  Code & $\alpha$(HOH) & d(OH) & E [a.u.] & $\alpha$(HNH) & d(NH) & E [a.u.] & $\alpha$(HCH) & d(CH) & E [a.u.] \\
\hline
Gamess         &  103.720 & 0.9656 & -76.3826 & 105.719 & 1.0183 & -56.5212 & 109.471 & 1.0921 & -40.4881\\
Uquantchem & 103.741 & 0.9656 & -76.3825 & 105.717 & 1.0183 & -56.5211& 109.470 & 1.0922 & -40.4878\\
\hline
\end{tabular}
\label{table:table2}
\end{table*}
\begin{table}[htb]
\centering
\caption{Experimental data for the equilibrium geometries for H$_{2}$O, NH$_{3}$ and CH$_{4}$   \cite{PBErev}.}
\begin{tabular}{c | c c}
\hline\hline
 & $\alpha$ & d  [\AA] \\
 \hline \hline
H$_{2}$O & 103.9 & 0.959\\
NH$_{3}$ & 106.0 & 1.012 \\
CH$_{4}$  & 109.5 & 1.086 \\
\hline
\end{tabular}
\label{table:table3}
\end{table}

\section{TEST-CALCULATIONS}
In order to test that the compilation of the UQUANTCHEM package have been successful, seven different test-calculations have been provided within the package. The input files of these
calculations are located in the subdirectories \texttt{RUN1,RUN2,$\ldots$,RUN7}  of  the \texttt{TESTS} -directory. Here follows a brief description of these tests:
\begin{itemize}
\item {\bf RUN1:} An unrestricted Hartree-Fock total energy calculation of a single water molecule.
\item {\bf RUN2:} A DFT total energy calculation using the revPBE functional  of a single water molecule.
\item {\bf RUN3:} A DFT total energy calculation and interatomic force calculation using the B3LYP functional  of a single water molecule.
\item {\bf RUN4:} A MP2 total energy calculation of a single water molecule.
\item {\bf RUN5:} A CISD  total energy calculation of a single Be atom.
\item {\bf RUN6:} A DQMC  total energy calculation of a single He atom, using a cc-PVTZ basis set.
\item {\bf RUN7:} An unrestricted Hartree-Fock structural relaxation of a single water molecule, using a STO-3G basis set.
\end{itemize}
In al tests listed above the  6-31G** basis  is used except in test calculations RUN6 and RUN7. To run the test calculations the user just have to execute the command:
\texttt{./runtests.pl} , in the root directory of the UQUANTCHEM code.

\section{INTERFACING WITH OTHER COMPUTATIONAL SOFTWARE}
The UQUANTCHEM software comes with a set of supporting utility scripts for generating  two-dimensional charge-density plots and three-dimensional iso-density plots with Matlab. See for example 
Figure \ref{fig:dqmc} showing a two-dimensional charge-density plot of a water molecule generated with the  \texttt{chargdens2DIM.m} Matlab script included in the software package.

\begin{figure}[tbp!]
\begin{center}
  \includegraphics*[angle=0,scale=0.3]{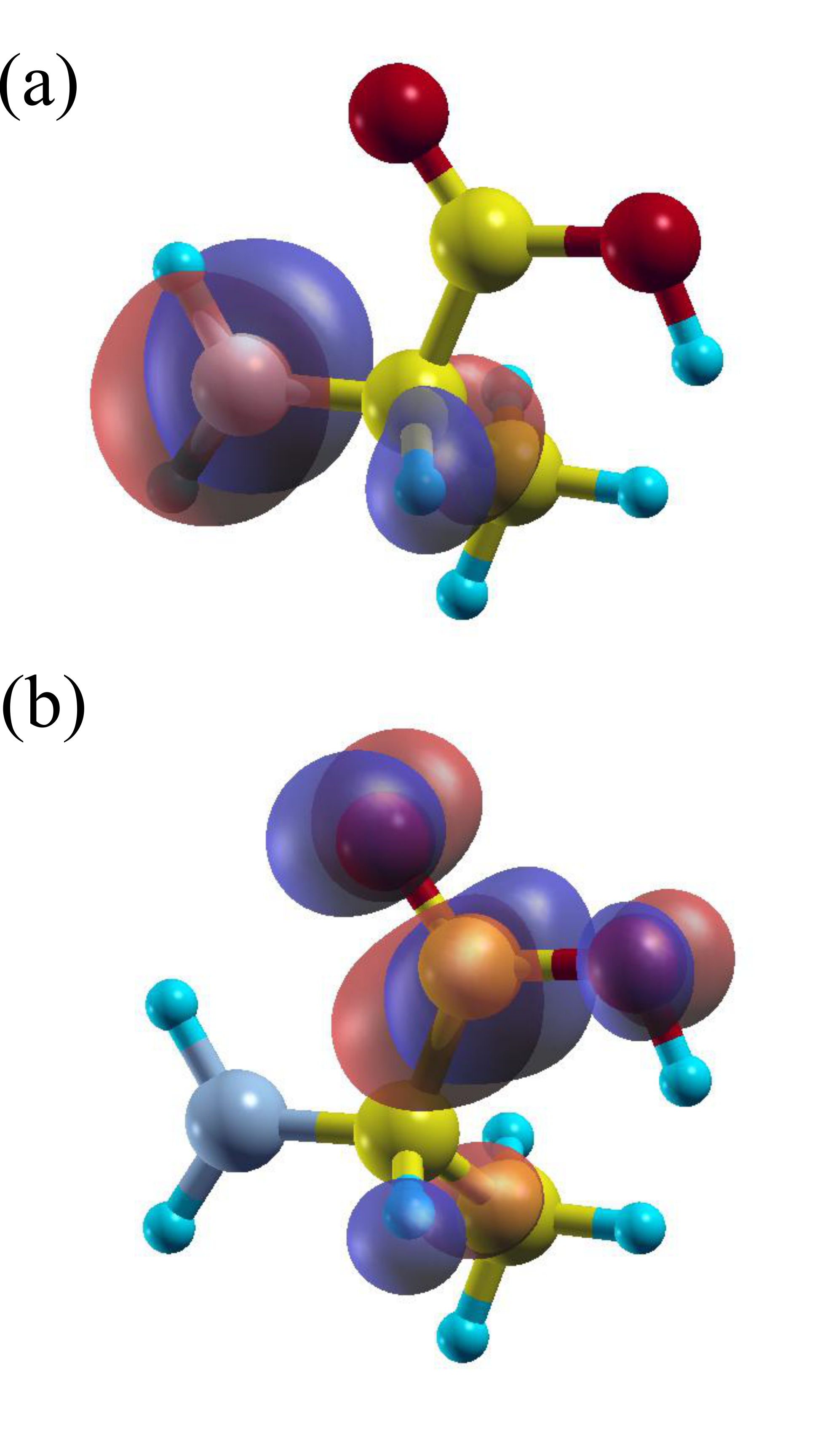}
  \caption{The highest occupied molecular orbital (HOMO) iso-surface, (a), and the lowest unoccupied molecular orbital (LUMO) iso-surface, (b), of the Alanine amino acid calculated with UQUANTCHEM using 
  a 6-31G** basis set together with the  B3LYP \cite{becke88,VWN,LYP,b3lyp}  functional. Here the rendering of the iso-surfaces was obtained by using the UQUANTCHEM output files \texttt{HOMO.xsf} and  
   \texttt{LUMO.xsf} as input to  the XCrySDen graphical software package \cite{xcrysden}.}
  \label{fig:homolumo}
  \end{center}
\end{figure}

The UQUANTCHEM has been adapted to work with the graphical software package XCrySDen  \cite{xcrysden}, by generating a number of files in the xsf-format. In Figure \ref{fig:homolumo}
the highest occupied molecular orbital (HOMO) iso-surface and the lowest unoccupied molecular orbital (LUMO) iso-surface of Alanine are shown. The 
rendering of the iso-surfaces was obtained by using the UQUANTCHEM output files \texttt{HOMO.xsf} and \texttt{LUMO.xsf} as input to  the xcrysden\cite{xcrysden} graphical software package.
%
\section{CONCLUSION}
In this work the recently developed quantum chemistry package UQUANTCHEM has been presented together with the results of some performance benchmark calculations. Here, even though  the scaling of the computational time 
with respect to system size, especially the MP2 and CISD implementations, leave room for much improvement, the MPI-implementation of the package compensates rather well for the limited performance in the serial version,  enabling the UQUANTCHEM 
package  to be utilized as a "proof of principle" platform  for new computational ideas in quantum chemistry, or to be utilized in standard quantum chemistry calculations of  molecules with $\leq$ 100 atoms.

\section*{Acknowledgements}
 Thanks to Professor Olle Eriksson for his patience and to Carl Caleman and Anders Niklasson for their inspiration.

The manual for the UQUANTCHEM software package can be obtained at:  \url{http://www.anst.uu.se/pesou087/UU-SITE/Webbplats_2/UQUANTCHEM_files/manual.pdf}





\bibliographystyle{elsarticle-num}







\end{document}